\documentclass[twocolumn,showpacs,aps,prl,superscriptaddress,nofootinbib]{revtex4}

\usepackage{graphicx}

\usepackage{dcolumn}

\usepackage{epsfig}

\RequirePackage{xspace}

\usepackage{relsize}
\def\babar{\mbox{\slshape B\kern-0.1em{\smaller A}\kern-0.1em
    B\kern-0.1em{\smaller A\kern-0.2em R}}}

\def\epem       {\ensuremath{e^+e^-}\xspace}

\def\qqbar {\ensuremath{q\overline q}\xspace}

\def\Kbar  {\kern 0.2em\overline{\kern -0.2em K}{}\xspace}

\def\Kz    {\ensuremath{K^0}\xspace}
\def\Kzb   {\ensuremath{\Kbar^0}\xspace}
\def\KzKzb {\ensuremath{\Kz \kern -0.16em \Kzb}\xspace}
\def\Kp    {\ensuremath{K^+}\xspace}
\def\Km    {\ensuremath{K^-}\xspace}

\def\KpKm  {\ensuremath{\Kp \kern -0.16em \Km}\xspace}

\def\Dbar    {\kern 0.2em\overline{\kern -0.2em D}{}\xspace}

\def\Dz      {\ensuremath{D^0}\xspace}
\def\Dzb     {\ensuremath{\Dbar^0}\xspace}
\def\DzDzb   {\ensuremath{\Dz {\kern -0.16em \Dzb}}\xspace}
\def\Dp      {\ensuremath{D^+}\xspace}
\def\Dm      {\ensuremath{D^-}\xspace}

\def\DpDm    {\ensuremath{\Dp {\kern -0.16em \Dm}}\xspace}
\def\Dstar   {\ensuremath{D^*}\xspace}

\def\Dstarp  {\ensuremath{D^{*+}}\xspace}
\def\Dstarm  {\ensuremath{D^{*-}}\xspace}

\def\Dstarmp {\ensuremath{D^{*\mp}}\xspace}

\def\Bbar    {\kern 0.18em\overline{\kern -0.18em B}{}\xspace}

\def\BB      {\ensuremath{B\Bbar}\xspace} 
\def\Bz      {\ensuremath{B^0}\xspace}
\def\Bzb     {\ensuremath{\Bbar^0}\xspace}
\def\BzBzb   {\ensuremath{\Bz {\kern -0.16em \Bzb}}\xspace}
\def\Bu      {\ensuremath{B^+}\xspace}
\def\Bub     {\ensuremath{B^-}\xspace}

\def\BpBm    {\ensuremath{\Bu {\kern -0.16em \Bub}}\xspace}

\def\BorBbar    {\kern 0.18em\optbar{\kern -0.18em B}{}\xspace}
\def\DorDbar    {\kern 0.18em\optbar{\kern -0.18em D}{}\xspace}
\def\KorKbar    {\kern 0.18em\optbar{\kern -0.18em K}{}\xspace}

\mathchardef\Upsilon="7107
\def\Y#1S{\ensuremath{\Upsilon{(#1S)}}\xspace}% no space before {...}!

\mathchardef\Deltares="7101
\mathchardef\Xi="7104
\mathchardef\Lambda="7103
\mathchardef\Sigma="7106
\mathchardef\Omega="710A

\def\Deltabar{\kern 0.25em\overline{\kern -0.25em \Deltares}{}\xspace}
\def\Lbar{\kern 0.2em\overline{\kern -0.2em\Lambda\kern 0.05em}\kern-0.05em{}\xspace}
\def\Sigbar{\kern 0.2em\overline{\kern -0.2em \Sigma}{}\xspace}
\def\Xibar{\kern 0.2em\overline{\kern -0.2em \Xi}{}\xspace}
\def\Obar{\kern 0.2em\overline{\kern -0.2em \Omega}{}\xspace}
\def\Nbar{\kern 0.2em\overline{\kern -0.2em N}{}\xspace}
\def\Xb{\kern 0.2em\overline{\kern -0.2em X}{}\xspace}

\newcommand{\tev}{\ensuremath{\mathrm{\,Te\kern -0.1em V}}\xspace}
\newcommand{\gev}{\ensuremath{\mathrm{\,Ge\kern -0.1em V}}\xspace}
\newcommand{\mev}{\ensuremath{\mathrm{\,Me\kern -0.1em V}}\xspace}
\newcommand{\kev}{\ensuremath{\mathrm{\,ke\kern -0.1em V}}\xspace}
\newcommand{\ev}{\ensuremath{\mathrm{\,e\kern -0.1em V}}\xspace}
\newcommand{\gevc}{\ensuremath{{\mathrm{\,Ge\kern -0.1em V\!/}c}}\xspace}
\newcommand{\mevc}{\ensuremath{{\mathrm{\,Me\kern -0.1em V\!/}c}}\xspace}
\newcommand{\gevcc}{\ensuremath{{\mathrm{\,Ge\kern -0.1em V\!/}c^2}}\xspace}
\newcommand{\mevcc}{\ensuremath{{\mathrm{\,Me\kern -0.1em V\!/}c^2}}\xspace}
\def\mus  {\ensuremath{\rm \,\mus}\xspace}

\def\mus        {\ensuremath{\,\mu{\rm s}}\xspace}    %% microsecond
\def\to                 {\ensuremath{\rightarrow}\xspace}

\def\pep2{PEP-II}

\def\gsim{{~\raise.15em\hbox{$>$}\kern-.85em
          \lower.35em\hbox{$\sim$}~}\xspace}
\def\lsim{{~\raise.15em\hbox{$<$}\kern-.85em
          \lower.35em\hbox{$\sim$}~}\xspace}

\def\CP                {\ensuremath{C\!P}\xspace}
\def\P       {\ensuremath{P}\xspace}
\def\T       {\ensuremath{T}\xspace}

\xspace

\newcommand{\jprlBase}       {Phys.\ Rev.\ Lett.\xspace}

\newcommand{\jprl}      [1]  {\jprlBase\ {\bf #1}}
\def\jetset74   {\mbox{\tt Jetset \hspace{-0.5em}7.\hspace{-0.2em}4}\xspace}
\def\resp#1{} %{ \small (#1) }}

\def\sinphi{\sin(2\beta + \gamma)}

\def\mmiss{m_{\rm miss}}

\def\Dstarp{\Dstar{^+}}
\def\Dstarm{\Dstar{^-}}

\def\btoc{b \rightarrow c \bar u  d}

\def\btodpipm{\Bz\rightarrow D^\mp\pi^\pm}
\def\btodstpipm{\Bz \rightarrow \Dstarmp\pi^\pm}
\def\btodstpi{\Bz \rightarrow \Dstarp\pi^-}

\def\btodstrhopm{\Bz\rightarrow \Dstarmp\rho^\pm}
\def\btodstrho{\Bz\rightarrow \Dstarp\rho^-}

\def\dt{\Delta t}
\def\dtErr{\sigma_{\dt}}

\def\dttrue{\dt_{\rm tr}}

\def\mc{MC}

\def\Brec{B_{\rm rec}}
\def\Btag{B_{\rm tag}}
\def\zrec{z_{\rm rec}}
\def\ztag{z_{\rm tag}}

\def\stag{s_{_{\rm t}}}
\def\smix{s_{_{\rm m}}}

\def\dstpi{{\Dstar\pi}}
\def\dstrho{{\Dstar\rho}}

\def\comb{{\it comb}}
\def\peak{{\it peak}}

\def\cont{{q\overline q}}

\def\P{{\cal P}}  % for total PDF
\def\T{{\cal T}}  % for Dt part
\def\M{{\cal M}}  % for mrec part
\def\F{{\cal F}}  % for fisher part
\def\R{{\cal R}}  % for resolution function
\def\r{r}

\newcommand{\BABARPubNumber}  {03/033}

\newcommand{\SLACPubNumber} {10204}

\def\figurebox#1#2#3{%
    \def\arg{#3}%
    \ifx\arg\empty
    {\hfill\vbox{\hsize#2\hrule\hbox to #2{\vrule\hfill\vbox to #1{\hsize#2\vfill}\vrule}\hrule}\hfill}%
    \else
    {\hfill\epsfbox{#3}\hfill}%
    \fi}

\begin{document}

%\preprint{\babar-PUB-02/013} 
%\preprint{SLAC-PUB-\SLACPubNumber} 

\begin{flushleft}
\babar-PUB-\BABARPubNumber\\
SLAC-PUB-\SLACPubNumber\\
%hep-ex/\LANLNumber\\[10mm]
\end{flushleft}

\title{\large \bf Measurement of 
Time-Dependent 
\boldmath \CP Asymmetries and Constraints on $\sin(2\beta+\gamma)$ with
Partial Reconstruction of $\btodstpipm$ Decays 
}

%%%%%%%%%%%%% \input authors_sep2003.tex
%% author list as of 01-Sep-2003 (603 authors)
%
\author{B.~Aubert}
\author{R.~Barate}
\author{D.~Boutigny}
\author{F.~Couderc}
\author{J.-M.~Gaillard}
\author{A.~Hicheur}
\author{Y.~Karyotakis}
\author{J.~P.~Lees}
\author{P.~Robbe}
\author{V.~Tisserand}
\author{A.~Zghiche}
\affiliation{Laboratoire de Physique des Particules, F-74941 Annecy-le-Vieux, France }
% ...
\collaboration{The \babar\ Collaboration}
\noaffiliation

%%%%%%%%%%%%% end of \input authors_sep2003.tex

\begin{abstract}
We present a measurement of time-dependent \CP-violating asymmetries in
decays of neutral $B$ mesons to the final states $\Dstarmp\pi^\pm$,
using approximately 
$82$ million $\BB$ events recorded by the \babar\ experiment
at the \pep2\ $\epem$ storage ring.  Events containing these decays are
selected with a partial reconstruction technique, in which only the
high-momentum $\pi^\pm$ from the $B$ decay and the low-momentum 
$\pi^\mp$ from the $\Dstarmp$ decay are
used. 
We measure the amplitude of the asymmetry to be  
$-0.063 \pm 0.024~(stat.) \pm 0.014~(syst.)$
and compute bounds on 
$|\sin(2 \beta + \gamma)|$.
\end{abstract}

% PACS, the Physics and Astronomy Classification Scheme.

\maketitle
\vskip .3 cm

%%%%%%%%%%%%%%%%%%%%%%%%%%%%%%%%%%%%%%%%%% body of paper

The Cabibbo-Kobayashi-Maskawa (CKM)
quark-mixing matrix~\cite{ref:km} gives an elegant explanation of \CP violation
and is under intense experimental investigation aimed at overconstraining 
its parameters. A crucial part of this program is 
the measurement of the 
angle $\gamma = \arg{\left(- V^{}_{ud} V_{ub}^\ast/ V^{}_{cd} V_{cb}^\ast\right)}$
of the unitarity triangle related to the CKM matrix.
The decay modes $\Bz \rightarrow {\Dstar}^{\mp}
h^{\pm}$, where $h = \pi,~\rho,~a_1$, have been
proposed for use in measurements of
$\sin(2\beta+\gamma)$~\cite{ref:book}, where $\beta = \arg{\left(-
V^{}_{cd} V_{cb}^\ast/ V^{}_{td} V_{tb}^\ast\right)}$.
In the Standard Model the decays 
$\Bz \to \Dstarp \pi^-$ and $\Bzb \to \Dstarp \pi^-$
proceed through the $\overline{b} \rightarrow \overline{u}  c  d $ and
$\btoc$ amplitudes $A_u$ and $A_c$. 
The relative weak phase between the two amplitudes 
in the usual Wolfenstein convention for the CKM matrix
is $\gamma$,
which, when combined with $\Bz \Bzb$ mixing, yields a weak phase
difference of $2\beta+\gamma$ between the interfering amplitudes.

The decay rate distribution for 
$B \to {\Dstar}^\pm\pi^\mp$ is
\begin{eqnarray}
\label{eq:pure-dt-pdf-B}
\P^\pm_\eta(\dt)
&=& {e^{-|\dt|/\tau} \over 4\tau} \times \\     
&&\left[1 \pm S^\zeta \sin(\Delta m \dt)
\mp \eta C \cos(\Delta m \dt)\right], \nonumber
\end{eqnarray}
where  
$\tau$ is the mean $\Bz$ lifetime,
$\Delta m$ is the $\Bz-\Bzb$ mixing 
frequency, and $\dt$
is the difference between the time
of the $B\to{\Dstar}^\pm\pi^\mp$ ($\Brec$)
decay and the decay of the other
$B$ ($\Btag$) in the event. The
upper (lower) sign in Eq.~\ref{eq:pure-dt-pdf-B}
indicates the flavor of the $\Btag$ as a $\Bz$ ($\Bzb$),
while $\eta = +1$ ($-1$) and $\zeta = +$ ($-$) for
the $\Brec$ final state ${\Dstar}^-\pi^+$ (${\Dstar}^+\pi^-$).
The parameters $C$ and $S^\pm$ are
%
%
%%%
\begin{equation}
C \equiv {1 - \r^2 \over 1 + \r^2}\, , \ \ \ \ 
S^\pm \equiv {2 \r \over 1 + \r^2}\, \sin(2 \beta + \gamma \pm \delta).
\label{eq:AandB}
\end{equation}
%%%%%%%%%%%%%%%%%%%%%%%
Here $\delta$ is the strong phase difference 
between $A_u$  and $A_c$ 
and 
$r = |A_u / A_c|$.
Since $A_u$ is doubly CKM-suppressed with respect
to $A_c$, one expects $r$ to be approximately $2\%$. 

In this Letter we report a study of \CP-violating asymmetries in
$\btodstpipm$ decays using the technique of partial reconstruction,
which allows us to analyse a large sample of signal events.
Additional information about the techniques used in this analysis 
appears in Ref.~\cite{ref:conf-paper,ref:dstpi-lifetime}.
%

%\section{PRINCIPLE OF THE MEASUREMENT}
%\label{sec:Principle}

%%%%%%%%%%%%%%%%%%%%%%%%%%%%%%%%%%%%%%%%%%%%%%%%%%%%%%%%%%%%%%%%%%%%
%\section{THE \babar\ DETECTOR AND DATASET}
%\label{sec:babar}

The data used in this analysis were recorded with the \babar\
detector at the \pep2\ storage ring, and consist of 76.4~fb$^{-1}$
collected on the $\Upsilon(4{\rm S})$ resonance (on-resonance
sample), and 7.6~fb$^{-1}$ collected at an $\epem$ center-of-mass (CM)
energy approximately 40~\mev below the resonance peak 
(off-resonance sample). Samples of simulated Monte Carlo (\mc) events
with an equivalent luminosity $3$ to $4$ times larger than the data are
analyzed through the same analysis chain.
The \babar\ detector is described in detail in Ref.~\cite{ref:babar}.

In the partial reconstruction of a $\btodstpipm$ candidate ($B_{rec}$), 
only the hard (high-momentum) pion track $\pi_h$ from the $B$ decay and the
soft (low-momentum) pion track $\pi_s$ from the decay
$D^{*-}\rightarrow \Dzb \pi_s^-$ are used.
This eliminates the efficiency loss associated with the
neutral $D$ meson reconstruction.
Applying kinematic constraints consistent with the signal decay mode, we
calculate the four-momentum of the $D$, obtaining its
flight direction to within a few
degrees and its invariant mass ($\mmiss$)~\cite{ref:dstpi-lifetime}.
Signal events peak in the $\mmiss$
%missing mass ($\mmiss$)~\cite{ref:dstpi-lifetime}
distribution at the nominal $\Dz$ mass with an r.m.s. of 3~\mevcc.

%%%%%%%%%%%%%%%%%%%%%%%%%%%%%%%%%%%%%%%%%%%%%%%%%%%%%%%%%%%%%%%%%%%%%%
%\subsection{Backgrounds}
%\label{sec:bgd}

In addition to $\btodstpipm$  events, the selected event sample  
contains the following kinds of events:
$\btodstrhopm$; 
$\BB$ background peaking in $\mmiss$,
  composed of pairs of tracks coming from
  the same $B$ meson, with the $\pi_s$ originating from a
  charged $\Dstar$ decay, excluding  $\btodstrhopm$ decays;
combinatoric $B$ background, defined as all remaining $\BB$ background events;
and continuum $\epem \rightarrow \qqbar$, 
where $q$
represents a $u$, $d$, $s$, or $c$ quark.
We apply an event-shape cut and a $\Dstar$ helicity-angle cut
to suppress combinatoric background. We  
reject $\pi_h$ candidates
that are identified as leptons or kaons.
All candidates must satisfy $1.81 < \mmiss <
1.88$~\gevcc. 
Multiple 
candidates are found in 5\% of the events. In these instances,
only the candidate with the $\mmiss$ value closest to $M_{\Dz}$ is used.

To perform this analysis, $\dt$ and the flavor of the $\Btag$ must be
determined. We measure $\dt$ using $\dt = (\zrec - \ztag) /
(\gamma\beta c)$, where $\zrec$ ($\ztag$) is the decay position of the
$\Brec$ ($\Btag$) along the beam axis ($z$) in the laboratory frame,
and the $e^+e^-$ boost parameter $\gamma\beta$ is
continuously calculated from the beam energies. 
To find $\zrec$ we fit the $\pi_h$ track with a beam
spot constraint in the plane perpendicular to the beams.
We obtain $\ztag$ 
from a beam-spot-constrained vertex fit
of all other tracks in the event, excluding all tracks within 
1~rad of the $D$ momentum in the CM frame.
The $\dt$ error $\dtErr$ is calculated from the results
of the $\zrec$ and $\ztag$ vertex fits.

We tag the flavor of the $\Btag$ using lepton or kaon
candidates.
The lepton CM momentum is required to be greater than 1.1~\gevc 
to suppress ``cascade'' leptons that originate from charm decays.
If several flavor-tagging tracks are present in either the lepton or kaon
tagging category,
the only track of that category used for tagging is the one with the
largest value of $\theta_T$, the CM angle between the track
momentum and the $D$ momentum.  The tagging lepton (kaon) must satisfy
$\cos \theta_T<0.75$ ($\cos \theta_T<0.50$), to minimize
the impact of tracks originating from the $D$ decay. If both a lepton and a
kaon satisfy this requirement, the event is tagged with the lepton only.

The analysis is carried out with a series of unbinned
maximum-likelihood fits, performed simultaneously on the on- and
off-resonance data samples and independently for the lepton-tagged and
kaon-tagged events.
The probability density function (PDF) depends on the variables
$\mmiss$, $\dt$, $\dtErr$, $F$, $\stag$, and $\smix$,
where 
$F$ is a Fisher discriminant formed from fifteen event-shape variables
that provide discrimination against continuum
events~\cite{ref:dstpi-lifetime},
$\stag = 1$ ($-1 $) when the $\Btag$ is identified as a $\Bz$ ($\Bzb$), 
and $\smix = 1$ ($-1 $) for ``unmixed'' (``mixed'') events.
An event is labeled unmixed if the $\pi_h$ is a 
$\pi^- (\pi^+)$ and the $\Btag$ is a $\Bz (\Bzb)$, and mixed
otherwise.

The PDF for on-resonance data is a sum over the PDFs of
the different event types,
$\P = \sum_{i}f_{i} \, \P_i$,
where the index $i = \{\dstpi, \dstrho, \peak, \comb, \cont\}$
indicates one of the event types described above, $f_i$ is the
relative fraction of events of type $i$ in the data sample, and $\P_i$
is the PDF for these events.
The PDF for off-resonance data is $\P_\cont$.
The parameter values for $\P_i$ are different for each event type, except
where indicated otherwise.
Each $\P_i$ is a product of the PDFs 
$\M_i(\mmiss)$, $\F_i(F)$, and $\T'_i(\dt, \dtErr, \stag, \smix)$.

The $\mmiss$ PDF $\M_i$ for each event type $i$ is the sum of a
bifurcated Gaussian and an ARGUS
function~\cite{ref:dstpi-lifetime}.
The Fisher PDF $\F_i$ is a bifurcated
Gaussian. The parameter values for $\F_\dstpi$,
$\F_\dstrho$, $\F_\peak$, and $\F_\comb$ are identical.

The $\dt$ PDF is a convolution,
$
\T'_i = 
        \int d\dttrue\, \T_i(\dttrue, \stag, \smix) \,
        \R_i(\dt - \dttrue, \dtErr),
$
%%%%%%%%%%%%%%%
where $\T_i$ is the distribution of the true decay-time difference
$\dttrue$ and $\R_i$ is a three-Gaussian resolution function that
accounts for detector resolution and effects such as systematic
offsets in the measured positions of vertices~\cite{ref:dstpi-lifetime}.

The PDF $\T_\dstpi(\dttrue, \stag, \smix)$ for signal events
corresponds to Eq.~\ref{eq:pure-dt-pdf-B} with $O(r^2)$ 
terms neglected, and with additional parameters that account
for imperfect flavor tagging:
\begin{eqnarray}
\T_\dstpi&=& E_{\tau}(\dttrue)       
        \left\{ \alpha (1+ \smix \kappa)
                + (1-\alpha) \left[\left(1-\stag\,\Delta\omega 
                     \right) 
                \right. \right. \nonumber\\
        &&+ \smix\,(1-2\omega)\, \cos(\Delta m
          \dttrue)                                    \nonumber\\ 
        &&  - \stag\,(1-2\omega)\,
%        S^{\stag \smix}
        S^{\xi}
                \,\sin(\Delta m \dttrue) 
        \left. \right] \left. \right\},
\label{eq:CP-pdf-alpharho}
\end{eqnarray}
where 
the mistag rate $\omega$ is the probability to
misidentify the flavor of the $\Btag$ averaged over $\Bz$ and $\Bzb$,
$\Delta \omega$ is the 
$\Bz$ mistag rate minus the $\Bzb$ mistag rate,
$\alpha$ is the probability that the tagging track is a
daughter of the signal $D$ meson,
$\kappa = 1-2 \rho$,
where $\rho$ is the probability that the daughter of the $D$ 
results in a mixed flavor
event, and $\xi=+(-)$ for $\stag \smix=+1(-1)$. 

The $\Btag$ may undergo a $b\rightarrow u \bar c d$ decay, and the kaon 
produced in the subsequent charm decay might be used for tagging.
This effect is not described by Eq.~\ref{eq:CP-pdf-alpharho}. 
To account for it, we use a different 
parameterization~\cite{ref:abc} for kaon tags,
in which the coefficient of the $\sin(\Delta m \dttrue)$ term
in Eq.~\ref{eq:CP-pdf-alpharho} is
\begin{equation}
- \left[ (1-2\omega) \, (\stag a + \smix c) 
        + \stag \smix b (1-\stag \Delta\omega) \right],
\label{eq:abc}
\end{equation}
where
$a\equiv 2 r\sin(2\beta+\gamma)\cos\delta$,
$b\equiv 2 r'\sin(2\beta+\gamma)\cos\delta'$,
and
$c\equiv 2\cos(2\beta+\gamma)(r\sin\delta - r'\sin\delta')$.
Here $r'$ ($\delta'$) is the effective magnitude ratio (strong phase
difference) between the $b\rightarrow u \overline c
d$ and $b\rightarrow c \overline u d$ amplitudes in the tag-side
decays.  This parameterization is good to first order in $r$ and $r'$.

%%%%%%%%%%%
%% D*rho %%
%%%%%%%%%%%
The PDF $\T'_\dstrho$ for $\btodstrhopm$ events is taken to be identical to
$\T'_\dstpi$, except for the 
\CP parameters ($S^\pm$, $a$, $b$, and $c$), 
which are set to 0.
The $\BB$ background PDFs $\T_\comb$ and $\T_\peak$ 
have the same functional form as
Eqs.~\ref{eq:CP-pdf-alpharho} and~\ref{eq:abc}, with independent parameter
values. 
The parameters of $\T'_\peak$ are determined from a fit to the \mc\
simulation sample. The \CP parameters of all the $\BB$ backgrounds are
set to $0$ and are later varied 
to evaluate systematic uncertainties.
The PDF $\T_\cont$ for the continuum background is
the sum of two components, one with a finite lifetime
and one with zero lifetime.

\begin{figure}[!htbp]
\begin{center}
        \includegraphics[width=0.49\textwidth]{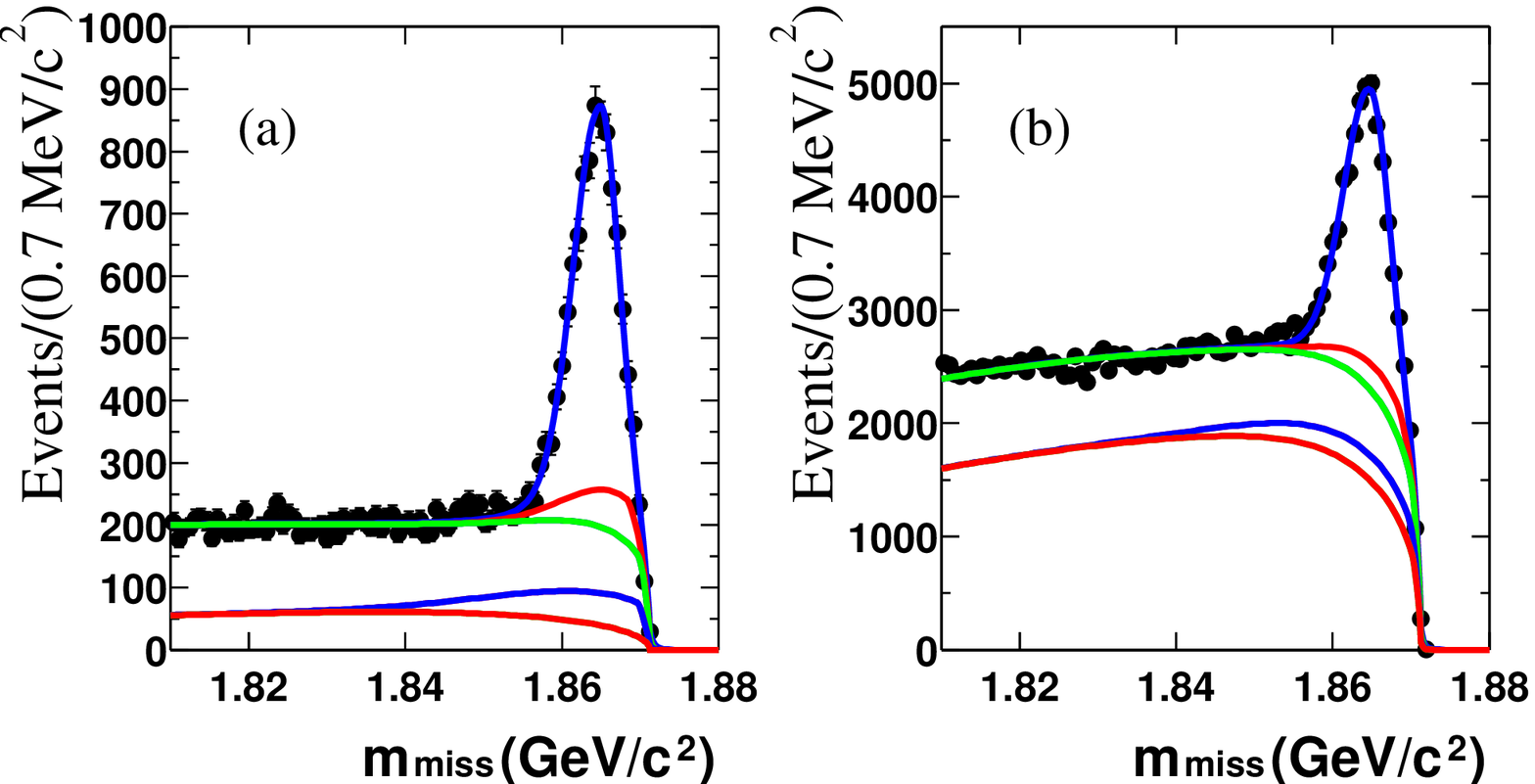}
\end{center}
\caption{The $\mmiss$ 
distributions for (a) lepton-tagged
and (b) kaon-tagged events.
The curves show, from bottom
to top, the cumulative contributions of continuum, peaking \BB, 
combinatoric \BB, $\btodstrho$,
and $\btodstpi$ events.
}
\label{fig:data_kin}
\end{figure}

The measurement of the \CP parameters proceeds in three steps:

1. The parameters of $\M_i$ and the
value of $f_\dstpi/(f_\dstpi + f_\dstrho)$ are obtained from the
\mc\ simulation with the branching fractions 
${\cal B}(\Bz \rightarrow \Dstarm\pi^+)$ and 
 ${\cal B}(\Bz \rightarrow \Dstarm\rho^+)$
from
Ref.~\cite{ref:pdg2002}.
Using these parameter values, 
we fit the data with 
$
\P_i = \M_i(\mmiss) \, \F_i(F),
$
to determine $f_{\cont}$, $f_{\comb}$, $f_\dstrho + f_\dstpi$, 
the parameters of $\M_\cont$, and the parameters of
$\F_i$ for both continuum and $\BB$ events. 
This fit yields $6400 \pm 130$ ($ 25160 \pm 320$) 
signal events for the
lepton- (kaon-) tagged sample.
Projections of the fit results are  shown in Fig.~\ref{fig:data_kin}. 
The fit is repeated to determine the relative signal yields above
and below the cut on $\cos \theta_T$, which determine
the values of $\alpha$ and $\rho$. We find 
$\alpha=(1.0 \pm 0.1)\%$ ($(5.6 \pm 0.2)\%$) for 
lepton- (kaon-) tagged events. 
%
%

%%%
%%%%%%%%%%%%%%%%%

2. We fit the events in the sideband $1.81 < \mmiss <
1.84$~\gevcc to obtain the parameters of
$\T'_{\comb}$ and $\T'_{\cont}$.

3. Using the parameter values obtained in
the previous steps (except $\T'_{\cont}$), we fit the data in the signal
region $1.845 < \mmiss < 1.880$~\gevcc, 
determining the parameters of $\T'_\dstpi$ and $\T'_\cont$.

We use the \mc\ samples to verify the entire analysis procedure, as
well as the validity of using the same non-\CP parameters in
$\T'_\dstrho$ and $\T'_\dstpi$ and of using the $\T'_\comb$ parameters
obtained from the sideband in the signal region.
For lepton-tagged events, we find a bias of $\mp 0.012$ in
$S^\pm$, due to the assumption that events tagged with direct
and cascade leptons are described by the same resolution
function. The results presented below are corrected for this bias.

\begin{figure}[!htbp]
\begin{center}
\includegraphics[width=0.49\textwidth]{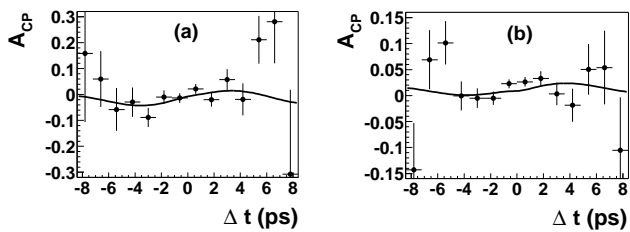}\\
\end{center}
\vspace*{-0.2cm}
\caption{The asymmetry $\mathcal{A}_{\CP}$ for (a) lepton- and (b) 
kaon-tagged events. The curves show the 
projection of the PDF from the unbinned fit.} 
\label{fig:asym}
\end{figure}

The signal-region fit (step 3) on the data determines the CP-parameters
$(S^+, S^-)$ for lepton tags and $(a,b,c)$ for kaon tags which are found to be
%The results of the signal-region fit (step 3) on the data are
%
   \def\dstpiA{{}}
%    change the above definition to \dstpi to get the subscript back
%
\begin{eqnarray}
S^+_\dstpiA &=& -0.078 \pm 0.052  \pm 0.021 , \nonumber\\
S^-_\dstpiA &=& -0.070 \pm 0.052  \pm 0.019 , \nonumber\\
a_\dstpiA &=& -0.054 \pm 0.032  \pm 0.017 , \nonumber\\
b_\dstpiA &=& -0.009 \pm 0.019  \pm 0.013 , \nonumber\\
c_\dstpiA &=& +0.005 \pm 0.031  \pm 0.017 ,
\label{eq:results}
\end{eqnarray}
where the first error is statistical and the second is systematic.  
The time-dependent, \CP-violating asymmetry
$
\mathcal{A}_{\CP} = 
(N_{\Bz_{\rm tag}} - N_{\Bzb_{\rm tag}}) /   
(N_{\Bz_{\rm tag}} + N_{\Bzb_{\rm tag}})
$
is shown in Fig.~\ref{fig:asym}.
In the absence of background and experimental effects,
$ \mathcal{A}_{\CP} = 2 r \sinphi \cos \delta \sin (\Delta m \Delta t )$.

%%%%%%%%%%%%%%%%%%%%%%
%%%%%%%%%%%%%%%%%%%%%%

%%%%%%%%%%%%%%%%%%%%%%%%%%%%%%%%%%%%%%%%%%%%%%%%%%%%%%%%%%%%%%%%%%
%\section{SYSTEMATIC STUDIES}
%\label{sec:Systematics}

The systematic uncertainties on the \CP-violation parameters 
are summarized in Table~\ref{tab:syst}. They include 
(1)~the statistical errors obtained from the fits of steps 1 and 2; 
(2)~uncertainties due to the unknown values of the \CP-violation 
parameters in the background,
the uncertainty in the ratio of branching fractions
${\cal B}(\Bz\to D^{*-} \pi)/ {\cal B}(\Bz\to D^{*-}
\rho)$, 
%~\cite{ref:pdg2002},
%
the modeling of $\T'_\peak$, and possible biases introduced
by the presence of background;
(3)~the uncertainty in the cascade lepton bias and possible biases
due to the $\tau$ and $\Delta m$ parameters;
(4)~uncertainties in the measurement of the beam spot position,
the detector $z$ length scale, and detector alignment; 
and (5)~the statistical error in the parameters determined from the MC sample.

\begin{table}[!htb]
\caption{The systematic uncertainties on the \CP-violation parameters.}
\begin{center}
\begin{tabular}{lccccc} 
\hline\hline
Source & \multicolumn{5}{c}{Error ($\times 10^{-3}$) in} \\ \cline{2-6}
      & $S^+$  & $S^-$  & $a$  & $b$  & $c$   \\
\hline
(1) Step 1 \& 2 statistics & 1.7 & 0.9 & 1.0 & 0.5 & 0.6 \\
(2) Backgrounds          &12.1 &10.0   &13.7  & 8.4  & 14.2 \\
(3) Fit procedure       & 6.6 & 5.3 & 5.2 & 1.7 & 0.8\\
(4) Detector effects     & 9.4 & 7.3 & 3.7 & 9.1     & 3.5 \\
(5) MC statistics        & 12.8 & 12.8 & 8.0  & 4.0      & 9.0 \\
\hline
Total                    & 21 & 19 & 17 & 13     & 17  \\
\hline\hline
\end{tabular}
\end{center}
\label{tab:syst}
\end{table}

%%%%%%%%%%%%%%%%%%%%%

Combining $a_\dstpiA$ and
$(S^+_\dstpiA + S^-_\dstpiA)/2$, accounting for correlated errors, we obtain
\begin{equation}
2 r \sinphi \cos \delta = -0.063 \pm 0.024 \pm 0.014.
\end{equation}
This measurement deviates from 0 by 2.3 standard deviations. 
It can be used 
to provide bounds 
on $|\sinphi|$~\cite{ref:Fleischer}.

We use two methods for interpreting our results in terms of
constraints on $|\sinphi|$. Both methods involve minimizing a $\chi^2$
function that is symmetric under the exchange $\sinphi \rightarrow
-\sinphi$, and applying the method of Ref.~\cite{ref:Feldman}.

\begin{figure}[htb]
\begin{center}
        \includegraphics[width=0.49\textwidth]{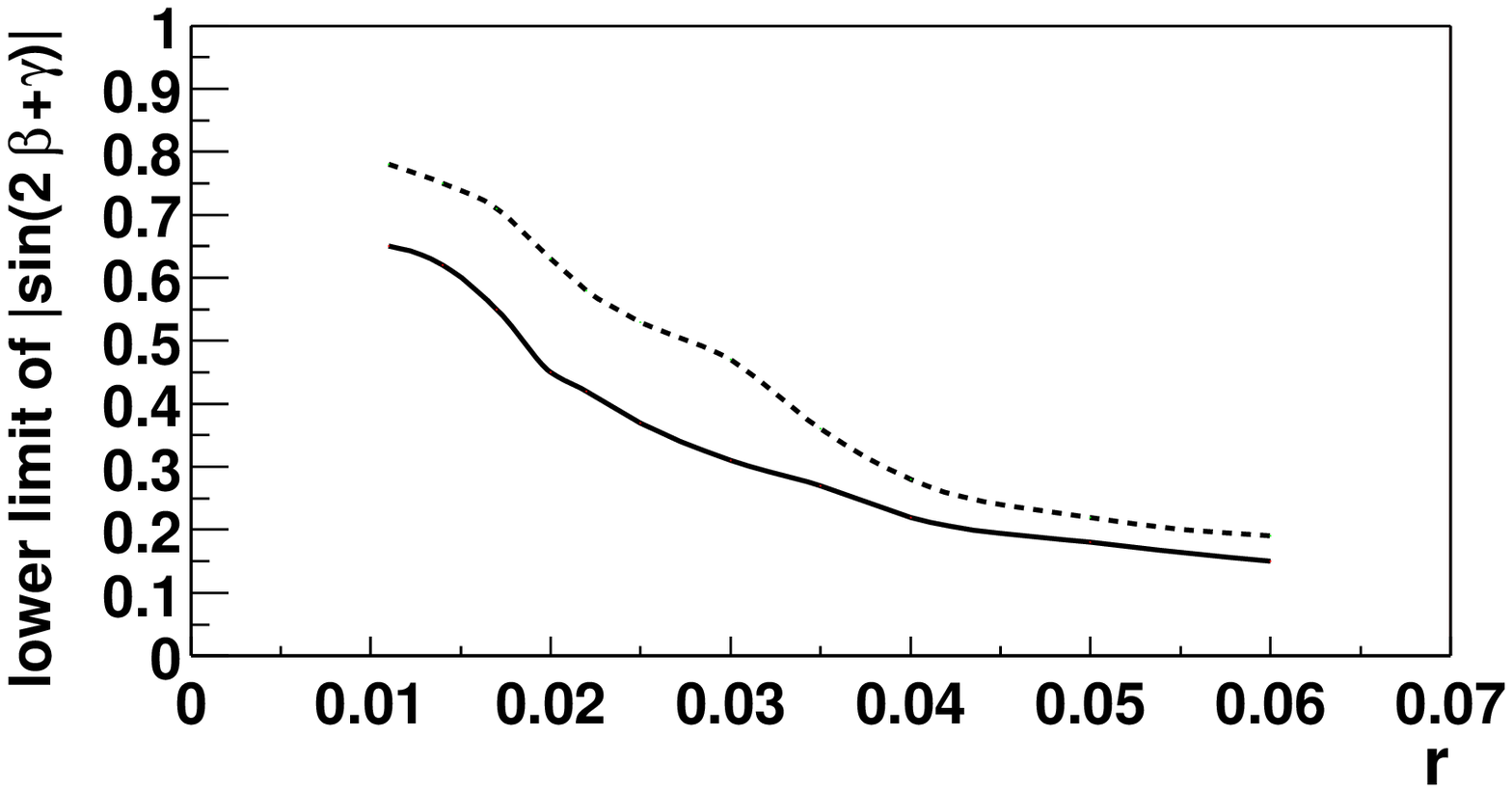}
\end{center}
\vspace*{-0.5cm}
\caption{
95\% CL lower limit on $|\sinphi|$ as a function of $r$. The solid curve 
corresponds to this analysis; the dashed curve includes the results of 
Ref.~\cite{ref:fullreco} for $\btodstpipm$ only.
}
\label{fig:limit-vs-r}
\end{figure}

In the first method we make no assumption regarding the value of $r$.
For different values of $r$ we minimize the function
$
\chi^2 = \sum_{j,k=1}^3 \Delta x_j V^{-1}_{jk} \Delta x_k ,
$
where $\Delta x_j$ is the difference between the result of our
measurement and the theoretical value of $S^+_\dstpiA$,
$S^-_\dstpiA$, and $a_\dstpiA$, and $V$ is the
measurement error matrix, which is nearly diagonal.
%
%%%%%%%%%%%%%%%%%%%%%%%%%%%%%%
The parameters determined by this fit are 
$\sin (2 \beta + \gamma)$, which is limited to lie
in the range $[-1, 1]$, and $\delta$.
We then generate many parameterized 
MC experiments with the
same sensitivity as reported here for different values of $\sinphi$ 
and with $\delta=0$, which yields the 
most conservative limits. 
The fraction of these experiments in which
$\chi^2(\sinphi) - \chi^2_{\rm min}$ is smaller than in the data is
interpreted as the confidence level
(CL) of the lower limit on $|\sinphi|$.
The resulting lower limit is shown as a function of $r$ in
Fig.~\ref{fig:limit-vs-r}.
This limit is always the more conservative of the two possibilities
implied by the ambiguity $|\sinphi|\leftrightarrow |\cos\delta|$.

The second method assumes that $r$ can be estimated from the
Cabibbo angle, the ratio of branching fractions ${\cal
B}(B^0\rightarrow {\Dstar}_s^{+} \pi^-) / {\cal B}(B^0\rightarrow
{\Dstar}^{-} \pi^+)$~\cite{ref:Dspi}, and the ratio of decay constants
$f_{\Dstar} / f_{\Dstar_s}$~\cite{ref:dec-const}, 
yielding 
$
r_0 = 0.017^{+0.005}_{-0.007}.
$
We attribute an additional
non-Gaussian 30\% relative error to the theoretical assumptions
involved in obtaining this value. 
We minimize
$
\tilde{\chi}^2 = \chi^2 + \Delta^2(r),
$
where $\Delta^2(r) = 0$ for $|r - r_0| / r_0 \le 0.3$ and is  
an offset quadratic function outside this range, corresponding to a 
$\chi^2$ contribution with the uncertainties on $r_0$ given above.
The parameters $\sin (2 \beta + \gamma)$, $\delta$, 
and $r$ are determined in this fit.
This method yields the limits 
$|\sinphi|> 0.87$ at 68\% CL and
$|\sinphi|> 0.56$ at 95\% CL.  

Combining this measurement with the BABAR results for fully reconstructed
$\btodstpipm$ and $\btodpipm$~\cite{ref:fullreco},
we find, using the second method, 
$|\sinphi|> 0.87$ at 68\% CL and
$|\sinphi|> 0.58$ at 95\% CL. 
The lower 
limit on $|\sinphi|$ obtained with the first method, 
including the results of Ref.~\cite{ref:fullreco} for $\btodstpipm$ only, is 
shown in Fig.~\ref{fig:limit-vs-r}.

To summarize, we have studied time-dependent \CP-violating asymmetries
in $\btodstpipm$ using partial reconstruction. The amplitude of the
\CP-violating asymmetry that we measure differs from zero by 2.3 standard
deviations.
We interpret our results as an $r$-dependent lower limit on $|\sinphi|$.

We are grateful for the excellent luminosity and machine conditions
provided by our \pep2\ colleagues, 
and for the substantial dedicated effort from
the computing organizations that support \babar.
The collaborating institutions wish to thank 
SLAC for its support and kind hospitality. 
This work is supported by
DOE
and NSF (USA),
NSERC (Canada),
IHEP (China),
CEA and
CNRS-IN2P3
(France),
BMBF and DFG
(Germany),
INFN (Italy),
FOM (The Netherlands),
NFR (Norway),
MIST (Russia), and
PPARC (United Kingdom). 
Individuals have received support from the 
A.~P.~Sloan Foundation, 
Research Corporation,
and Alexander von Humboldt Foundation.

\end{document}